# HyTopo：A High-Performance Hierarchical Algorithm for Mapping the Structural Evolution and Transition Networks of Gas Hydrates

Zherui Chen [a, b], Shuquan Huang [a], Ke Xu [c, d*], Zhanbo Li [a], Xinyu Song[a], Haijun Wang [e, f], Dingyuan Tang [a], Qingxia Liu [a, g*], Yajun Deng [a*]

[a] Future Technology School, Shenzhen Technology University, Shenzhen 518118, P. R. China

[b] College of Applied Sciences, Shenzhen University, Shenzhen 518060, P. R. China

[c] College of Physical Science and Technology, Bohai University, Jinzhou 121013, P. R. China

[d] Department of Electronic Engineering and Materials Science and Technology Research Center, The Chinese University of Hong Kong, Hong Kong 999077, P. R. China

[e] School of Energy Science and Engineering, Harbin Institute of Technology, Harbin 150040, P. R. China

[f] Zhengzhou Advanced Research Institute of Harbin Institute of Technology, Zhengzhou 450000, P. R. China

[g] Department of Chemical and Materials Engineering, University of Alberta, Edmonton, AB T6G 1H9, Canada

*Corresponding author: kickhsu@gmail.com(Ke Xu), dengyajun@sztu.edu.cn (Yajun Deng), qingxia2@ualberta.ca (Qingxia Liu)

**Abstract:** Resolving the formation and dissociation mechanisms of hydrate requires identifying not only hydrate-like water, but also the specific cage topologies by which disordered liquid water reorganizes into, or escapes from, crystalline hydrate networks. Here we introduce HyTopo, a cage-topology-centered analysis framework that combines a hydrate order parameter, hierarchical ring and cage recognition, and cage-transition network analysis. Based on local orientational correlations, HyTopo classifies water molecules into liquid-like, precursor-like, hydrate-like, and ice-like environments, and uses this information to guide efficient topological searches for crystalline hydrate cages and defective intermediates. Applications to methane hydrate growth, homogeneous nucleation, and depressurization-induced dissociation reveal that hydrate evolution is governed by heterogeneous and pathway-dependent cage dynamics. During hydrate growth, EA-WPU suppresses the overall proliferation of cage-like structures while altering the balance between defective cages and crystalline sI cages. During nucleation, incomplete cage motifs dominate the early pathway and provide a structural reservoir from which a small fraction of stable $5^{12}$ and $5^{12}6^{2}$ cages. Hydrate dissociation proceeds through interfacial collapse, bulk decomposition, and dissolved-gas exsolution, with cage lifetimes and transition pathways strongly dependent on cage topology. These results show that HyTopo provides a physically interpretable and topology-resolved route to connect local water ordering, cage formation, structural defects, and hydrate phase transitions in complex molecular simulations.

# 1. Introduction

Water is deceptively simple in composition yet remarkably diverse in structure[1-5]. From crystalline ice polymorphs to confined liquids, amorphous phases, and hydrogen-bonded host–guest frameworks, water continually challenges our understanding of how a small triatomic molecule organizes into complex collective architectures[6-10]. Among these forms, clathrate hydrates occupy a particularly fascinating position. In these ice-like inclusion compounds, water molecules assemble into three-dimensional hydrogen-bonded cages that encapsulate guest species such as methane, carbon dioxide, hydrogen, and larger amphiphilic molecules[11-13]. This combination of molecular simplicity and topological richness makes clathrate hydrates a central topic in water science, energy technology, environmental chemistry, and materials physics[14-17].

The importance of clathrate hydrates extends far beyond their structural elegance. Natural methane hydrates represent a vast potential energy reservoir[18-20], whereas hydrate formation in pipelines remains a major flow-assurance challenge in the oil and gas industries[21-23]. Carbon dioxide and hydrogen hydrates are closely related to carbon sequestration and hydrogen storage, respectively[24-29]. Hydrate-like ordering is also relevant to broader problems in aqueous self-assembly, including hydrophobic hydration, “iceberg” ordering around nonpolar groups, antifreeze protein activity, and cryopreservation[30-32]. More generally, clathrate hydrates provide molecular realizations of Frank–Kasper-type polyhedral frameworks, linking water science to a wider family of tetrahedral and cage-forming materials[33-36]. Understanding how water selects, distorts, connects, and transforms these polyhedral motifs is therefore essential not only for water science, but also for the broader physics of network-forming matter[37-39].

Despite decades of experimental, theoretical, and computational effort, the molecular mechanism of hydrate nucleation remains incompletely understood, largely because the decisive early structures are transient, disordered, and topologically heterogeneous[40-42], and because crystal nucleation in general is intrinsically stochastic[43-46], with rare events and pathway-to-pathway variability obscuring deterministic structural descriptions. Hydrate formation is now widely regarded as a multistep process in which guest molecules first aggregate into solvent-separated clusters and then promote the reorganization of surrounding water into amorphous hydrate-like intermediates composed of standard and non-standard polyhedral cages[47-50]. Identifying these early cages is essential for resolving polymorph selection, the maturation of amorphous precursors into crystalline domains, and the role of guest molecules in shaping the nucleation free-energy landscape. However, existing analysis methods remain limited. Local order parameters, such as Steinhardt-type descriptors and CHILL+-like[51] hydrogen-bond classifications, can monitor crystallization and assign local phase environments but cannot explicitly resolve cage topology, ring composition, or cage connectivity. Although topological algorithms can directly identify water rings and polyhedra, fully general searches are often prohibitively expensive for large trajectories. Some cage recognition programs, such as GRADE[52], HTR[53], improve efficiency through a hierarchical this cage-recognition strategy, but it primarily targets canonical cages such as $5^{12}$ and $5^{12}6^{2}$. As a result, existing programs still struggle to balance three competing requirements: recognizing incomplete and nonstandard cages, tracking their topological transformations, and maintaining the computational efficiency required for large molecular dynamics trajectories.

A further limitation is that most analysis tools treat trajectories as collections of independent snapshots[51-53]. They identify cage populations at each frame, but rarely reconstruct how individual cages emerge, fluctuate, transform, merge, or disappear. Hydrate nucleation, however, is inherently dynamic: guest-rich liquid clusters evolve into amorphous cage aggregates and eventually into crystalline hydrate domains through continuous structural rearrangements[54-56]. A mechanistic description therefore requires a transition-network framework that connects cage identities across time while preserving both local phase information and global cage topology[57, 58]. Such a framework must also handle periodic boundary conditions robustly, since extended hydrogen-bond networks and cage clusters often span simulation boundaries.

Here, we introduce HyTopo, a high-performance computational suite for mapping the structural evolution and transition networks of gas hydrates. HyTopo integrates a Numba-accelerated hierarchical topological search algorithm with a Hydrate Order Parameter classification kernel, enabling simultaneous identification of local hydrate-like environments and explicit recognition of standard, non-standard, distorted, and semi-cage topologies. By coupling local order classification with global ring-and-cage search, HyTopo improves the physical reliability of cage assignment while preserving the topological sensitivity required to capture amorphous hydrate intermediates. In addition, an automated unwrapping procedure is incorporated to handle periodic boundary conditions, ensuring consistent reconstruction of extended hydrogen-bond networks and cage clusters across simulation boundaries.

Beyond static cage recognition, HyTopo is designed to reconstruct the temporal evolution of hydrate cages. The algorithm tracks cage identities across molecular dynamics trajectories and builds transition networks that describe cage formation, annihilation, and interconversion. Its coordination profiles and ring-size distributions agree with atomistic benchmarks, while the accelerated implementation enables efficient analysis of large trajectory datasets. By revealing how distinct cage motifs emerge, transform, and contribute to hydrate-like order, HyTopo shifts hydrate analysis from static cage counting toward a dynamic description of evolving polyhedral networks, providing a unified framework for studying hydrate self-assembly, decomposition, and polymorphic transformation.

# 2. Theoretical Foundations

## 2.1. Hydrate Order Parameter (HOP) and Local Phase Classification

A robust identification of local water environments in HyTopo is established using a probabilistic phase-classification framework, based on which a new hydrate order parameter (HOP) was proposed. Specifically, water molecules in the system were classified into six categories: hydrate (HYD), cubic ice (Ic), hexagonal ice (Ih), precursors or amorphous hydrate (PRE), and liquid water (LIQ). A schematic illustration of the algorithm is shown in Figure 1.

Figure 1. HOP algorithm schematic diagram

This scheme extends the correlation of local bond orientational order parameters, originally proposed by ten Wolde et al. and refined in the CHILL+ algorithm[51]. In this framework, the local environment of a water oxygen atom $i$, is described by a $(2l+1)$- dimensional complex vector $q_l(i)$, defined by the projection of its four nearest neighbors onto spherical harmonics:

$$\mathbf{q}_{3m}(i)=\frac{1}{4}\sum_{j=1}^{4}Y_{3m}(\theta_{ij},\phi_{ij}) \tag{1}$$

where $\theta_{ij}$ and $\phi_{ij}$ are the polar and azimuthal angles of the vector connecting atom $i$ to its neighbor $j$. $Y_{3m}$ denotes the spherical harmonics of rank $l=3$. The spatial coherence between a pair of hydrogen-bonded water molecules $i$ and $j$ is evaluated via the normalized dot product of their respective order parameter vectors:

$$c(i,j)=\frac{\sum_{m=-3}^{3}q_{3m}(i)\cdot q_{3m}^{*}(j)}{\left[\sum_{m=-3}^{3}|q_{3m}(i)|^{2}\right]^{1/2}\left[\sum_{m=-3}^{3}|q_{3m}(j)|^{2}\right]^{1/2}} \tag{2}$$

Following the geometric criteria of CHILL+, a bond is classified as staggered (S) if $c(i,j)\leqslant -0.8$ and eclipsed (E) if $-0.35\leqslant c(i,j)\leqslant 0.25$.

Although CHILL+ is effective for identifying bulk ice under strong supercooling, its hard-threshold classification becomes fragile during spontaneous hydrate nucleation. Because phase labels are assigned from discrete integer counts of staggered and eclipsed bonds, small thermal distortions near interfaces or early nuclei can cause rapid label blinking between liquid-like, crystalline, and interfacial states, obscuring nucleation kinetics. Moreover, CHILL+ treats each molecule mainly through its immediate bonds and lacks the collective context of the surrounding hydrogen-bond network, making amorphous hydrate precursors prone to misclassification as liquid or generic interfacial water. To address these issues, HyTopo improves tolerance to thermal fluctuations and incorporates local environmental order, thereby enhancing the recognition of hydrate precursors.

To dampen high-frequency thermal noise without sacrificing spatial resolution, HyTopo implements a continuous local environment averaging scheme using Compressed Sparse Row (CSR) graph representations. For each molecule $i$, we define smoothed descriptors ($\chi_{\mathrm{env}}$) for the staggered count (S), eclipsed count (E), and local tetrahedrality ($q_{tet}$) as:

$$\chi_{\mathrm{env}}(i) = (1-\lambda)\,\chi(i) + \lambda\,\frac{1}{N_i}\sum_{j\in \mathrm{Adj}(i)} \chi(j) \tag{3}$$

Where $\chi \in \{S,\ E,\ q_{tet}\}$, $\mathrm{Adj}(i)$ represents the set of neighbors within the O-O cutoff, $N_i$ is the coordination number, and $\lambda$ is a mixing coefficient (typically 0.15). This spatial low-pass filter ensures that the resulting phase assignment reflects the collective order of the local domain rather than instantaneous single-molecule fluctuations.

Leveraging the refined structural descriptors derived in Eq (3), HyTopo transcends the limitations of binary integer counting by mapping the smoothed local field onto a continuous phase probability spectrum. In contrast to the rigid thresholding of CHILL+, which necessitates exact bond counts (e.g., E=4 for hydrates), our formalism utilizes a "soft-gating" approach where each water molecule is characterized by its propensity $\phi$ to exist in a specific phase. This transformation is achieved through a combination of sigmoidal (Fermi-Dirac) kernels and Gaussian functions, which naturally account for the structural overlaps and lattice distortions prevalent during the stochastic nucleation process. The propensity for the crystalline hydrate state ($\phi_{\mathrm{hyd}}$) and the transient amorphous precursor state ($\phi_{\mathrm{pre}}$) are constructed as multi-variable product functions:

$$\phi_{\mathrm{pre}}(i) = q_{\mathrm{env}}(i) \cdot f_{\mathrm{coord}}(i) \cdot f_{\mathrm{step}}(q_{\mathrm{env}}; q_{\mathrm{pre}}^{\mathrm{min}}) \cdot f_{\mathrm{gate}}(E_{\mathrm{env}}; E_{\mathrm{pre}}^{\mathrm{low}}, E_{\mathrm{pre}}^{\mathrm{high}}) \cdot f_{\mathrm{less}}(S_{\mathrm{env}}; S_{\mathrm{pre}}^{\mathrm{max}}) \quad (4)$$

$$\phi_{\mathrm{hyd}}(i) = q_{\mathrm{env}}(i) \cdot f_{\mathrm{coord}}(i) \cdot \exp\left[-\frac{(E_{\mathrm{env}}(i) - E_{\mathrm{hyd}}^{\mathrm{center}})^2}{2\sigma_E^2}\right] \cdot f_{\mathrm{less}}(S_{\mathrm{env}}; S_{\mathrm{hyd}}^{\mathrm{max}}) \cdot f_{\mathrm{step}}(q_{\mathrm{env}}; q_{\mathrm{hyd}}^{\mathrm{min}}) \quad (5)$$

Here, the gating kernels are defined to provide a smooth transition across phase boundaries: $f_{\mathrm{step}}(x;\ \theta,\ \omega) = [1 + \exp(-(x-\theta)/\omega)]^{-1}$ represents a structural activation gate, while $f_{less}(x;\ \theta,\ \omega) = [1 + \exp((x-\theta)/\omega)]^{-1}$ serves as a penalty gate for excessive bond types. The term $f_{\mathrm{coord}}(i)$ is a coordination gate ensuring the molecule resides within the physical range of a tetrahedral network.

The liquid-phase propensity is formally treated as the structural complement of the ordered states, defined as $\phi_{liq}(i) = \left[1 + \kappa \sum \phi_{solid}(i)\right]^{-1}$. $\kappa$ scales the sensitivity to solid-like ordering. To yield a physically interpretable phase assignment, these propensities are normalized into a partitioned probability spectrum:

$$P_\alpha(i) = \frac{\phi_\alpha(i)}{\sum_\beta \phi_\beta(i)} \qquad \alpha,\ \beta \in \{liq,\ Ih,\ Ic,\ hyd,\ pre\} \quad (6)$$

This continuous formulation allows HyTopo to capture the "mixed-phase" character of water molecules at the critical nucleus interface, providing a quantitative metric for the amorphous-to-crystalline transition that is fundamentally inaccessible to discrete counting algorithms.

In conventional workflows, cage search algorithms are applied to the entire hydrogen-bond network, incurring significant computational overhead in bulk liquid regions. HyTopo overcomes this by formally partitioning the system based on phase propensities.

Since water molecules residing in highly confident liquid states cannot physically participate in stable hydrate cages, HyTopo constructs a dynamic candidate mask:

$$\mathcal{M}(i)=\begin{cases}1, & \text{if } P_{\mathrm{hyd}}(i)+P_{\mathrm{pre}}(i)\geq\theta_{\mathrm{hyd}} \text{ or } (P_{\mathrm{Ih}}(i)+P_{\mathrm{Ic}}(i)\geq\theta_{\mathrm{ice}} \text{ and Include Ice}) \\ 0, & \text{if } P_{\mathrm{liq}}(i)\geq\theta_{\mathrm{liq_exclude}}\end{cases} \tag{7}$$

This candidate mask $\mathcal{M}(i)$ is used to prune the global hydrogen-bond graph prior to ring and cage perception. By isolating the topological search to a localized, structurally relevant subset of water molecules, HyTopo bypasses graph traversal in bulk liquid regions. This "gated search" strategy reduces the complexity of the graph and significantly improves the speed of searching for hydrate cage structures in the early stages of nucleation.

**2.2. Hierarchical Ring Perception and Geometric Filtration**

Following the probabilistic local phase classification, the active water molecules are mapped onto a pruned undirected subgraph $G_p=(V_p,\ E_p)$ to initiate the topological search. This graph-pruning step, which excludes bulk liquid-like molecules via the candidate mask $\mathcal{M}(i)$, restricts the search space to the nascent hydrate front, thereby bypassing the computationally expensive graph traversals inherent in traditional algorithms. Within this localized subgraph, hydrogen bonds are defined using a spatial oxygen–oxygen cutoff $R_{cut}=3.5\ \mathring{A}$ and a donor–acceptor angle limit $\angle HOO\leqslant 30°$。HyTopo then employs a high-performance, Numba-compiled depth-first search (DFS) kernel to enumerate all simple, chordless rings of sizes $n_r\in\{4,\ 5,\ 6\}$, which represent the fundamental polygonal faces of clathrate polyhedra. To suppress the combinatorial explosion of fluctuating loops at highly disordered liquid–hydrate interfaces, each candidate ring must survive a rigorous two-stage geometric filter.

The first stage of this filtration relies on convexity pruning, where topological path lengths within the ring are constrained. Specifically, a candidate loop of size $n_r = 5 \ or \ 6$ is discarded if the minimum image distance between any non-adjacent vertices falls below a geometrically defined threshold:

$$\begin{aligned} d(k,\ k+2) &\geqslant (1.6 \times R_{cut}) - \delta_1 \ \ (for\ n_r = 5) \\ d(k,\ k+3) &\geqslant (2.0 \times R_{cut}) - \delta_2 \ \ (for\ n_r = 6) \end{aligned} \tag{8}$$

where the tolerance parameters are set to $\delta_1 = 1.8\ \mathring{A}$ and $\delta_2 = 2.6\ \mathring{A}$ default. Rings satisfying this convexity criterion are then subjected to a singular value decomposition (SVD) of their unwrapped coordinates to assess planarity. By solving the eigenvalue problem for the centered coordinate matrix $A = r_k^{unwrapped} - r_{center}$, we obtain the surface normal $\hat{n}$ associated with the minimum singular value. The planarity deviation is defined as:

$$\Delta_{planarity} = \max_k |A \cdot \hat{n}| \tag{9}$$

and any loop exceeding a threshold of 1.5° is immediately eliminated. This planarity gate ensures that the normal vector of each surviving face is physically well-defined, providing a reliable basis for subsequent directional coverage and topological matching.

**2.3. Cup Assembly and Cage Construction**

The validated polygonal faces are subsequently assembled into localized, semi-enclosed manifolds termed "cups." A valid cup is topologically defined by a central bottom ring $R_{bottom}$ of size $n_c$ and a surrounding closed belt of $n_c$ lateral rings, where each lateral ring shares exactly one edge with the bottom ring and one edge with each of its two adjacent lateral neighbors. By ensuring that all lateral rings propagate on the same side of the bottom ring's plane, HyTopo systematically avoids self-intersecting "pseudo-cup" configurations. For symmetric, bulk crystalline phases, the reconstruction of standard edge-saturated cages (SECs) is achieved through a direct cup-overlapping mechanism. Under this scheme, pairs of complementary "twin cups" that propagate in opposite directions from a shared bottom ring are matched. For example, the ideal overlapping of two $5^6$ cups yields a standard $5^{12}$ cage, whereas the pairing of $6^1 5^6$ cups reconstructs the $5^{12}6^2$ cage.

To resolve the highly distorted, non-standard, or partially open cages that dominate the amorphous intermediate phase, HyTopo bypasses the rigid constraints of symmetric cup overlapping by implementing a heuristic beam-search algorithm. Starting from an initial seed ring, the cage is grown iteratively by adding neighboring rings to a candidate set $\Omega = \{r_1, r_2, \ldots, r_k\}$. The trajectory of this topological growth is guided by a heuristic scoring function:

$$S(\Omega) = \Sigma_{r \in \Omega} Score(r) + \alpha \cdot C(\Omega) \tag{10}$$

which balances the SVD-based planarity of individual faces with a global closure metric $C(\Omega) = 2E_{shared} / E_{total}$. To maintain computational tractability over large search spaces, a predefined beam width B (typically 64) and candidate limit (typically 20) are enforced to prune low-scoring search paths at each iteration. Upon reaching the target face count for a specific polyhedron, the topological closure of the candidate assembly $\Omega$ is verified using the Euler characteristic $\chi = V_{nodes} - E_{edges} + F_{faces}$. A candidate is identified as a complete, physically closed cage only if $\chi = 2$, the graph is fully connected, and all internal edges are shared by exactly two faces. For highly deformed clusters where complete closure cannot be achieved, the search automatically falls back to classify partial cages based on their boundary edge counts and fractional directional coverage. Figure 2 is a schematic diagram of cage structure search.

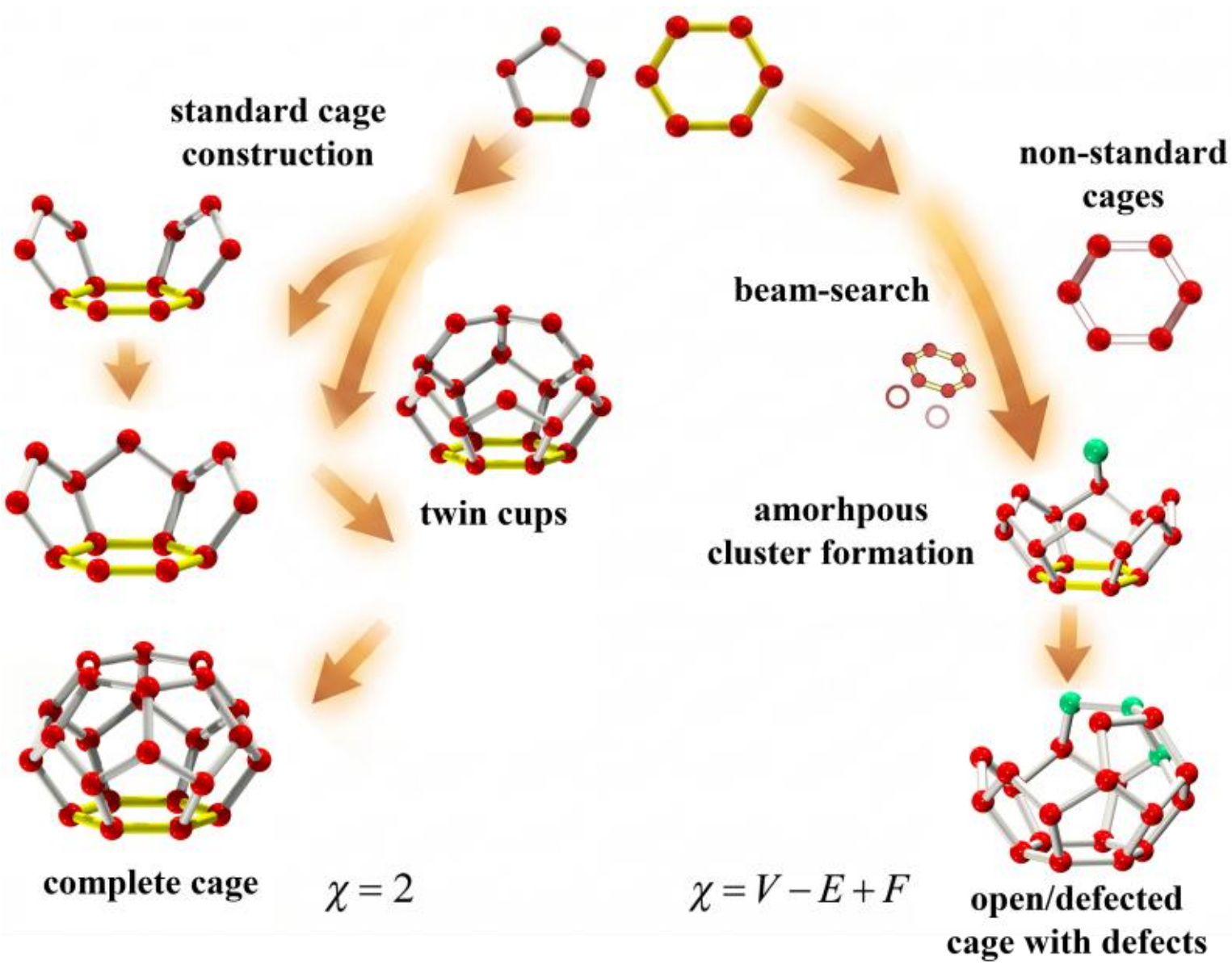

Figure 2. Schematic illustration of the HyTopo algorithm for hydrate cage structure recognition

# 3. Results and Discussion

## 3.1. Identification of Hydrate Growth Process

Hydrate formation in subsea oil and gas pipelines is a major flow assurance challenge, as the low-temperature and high-pressure operating conditions strongly promote hydrate growth and can ultimately lead to pipeline blockage. As a representative application of HyTopo, we investigated methane hydrate growth in the absence and presence of EA-WPU, a newly developed hydrate inhibitor reported in our previous work[59, 60]. Figure 3 compares the hydrate order parameter (HOP) calculated by HyTopo and the corresponding molecular configurations during a 500 ns hydrate growth process at 260 K and 50 MPa. In the uninhibited system, the HOP increased continuously from 0.31 to approximately 0.80 within the first 300 ns, indicating rapid bulk hydrate growth. In contrast, the system containing EA-WPU exhibited a markedly delayed growth process, with HOP values remaining substantially lower than those of the uninhibited system, particularly during the early and intermediate stages. During the first 300 ns, most water molecules in the EA-WPU-containing system remained in liquid-like or precursor-like states, whereas a well-defined hydrate structure had already formed in the corresponding uninhibited system. The inhibitor-rich interfacial layer appears to disrupt the ordering of water molecules around methane, thereby suppressing the formation of a continuous hydrate network.

Importantly, compared with the conventional $F_4$ order parameter, HOP provides higher structural resolution. $F_4$ is based on the dihedral angle formed by the outermost hydrogen atoms of two neighboring water molecules and the oxygen atom, and it can distinguish broadly between ordered and disordered water environments; however, its ability to separate crystalline hydrates, amorphous hydrates or hydrate precursors, and liquid water is limited. In our previous study, the curves for the inhibited and uninhibited systems were nearly indistinguishable over the 400–500 ns time interval[59, 60]. By contrast, HOP enables more sensitive discrimination among liquid water, hydrate, and precursor states. This enhanced resolution is particularly valuable for inhibitor screening, because effective inhibitors often function by stabilizing transient precursor or amorphous interfacial states rather than completely suppressing local water ordering. Therefore, HyTopo provides a more discriminative metric for evaluating hydrate growth inhibition and for identifying high-performance kinetic hydrate inhibitors such as EA-WPU.

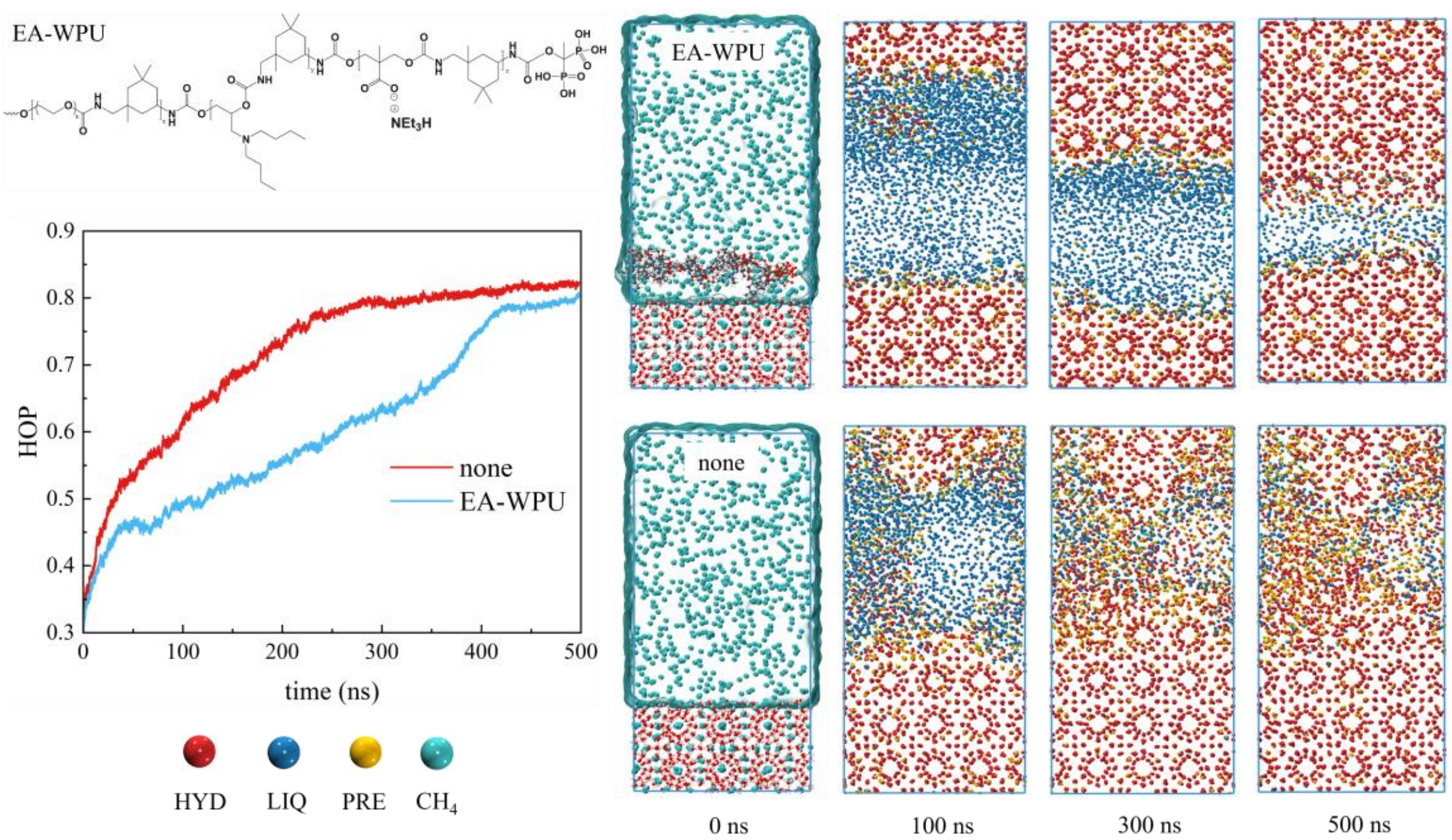

Figure 3. Hydrate order parameters (HOP) and corresponding molecular configurations during hydrate growth process

The HyTopo-resolved cage statistics reveal a nontrivial effect of EA-WPU on hydrate growth that cannot be captured by a single global order parameter. As shown in Figure 4(a), the population of canonical sI hydrate cages, defined as the sum of $5^{12}$ and $5^{12}6^{2}$ cages, is consistently lower in the EA-WPU system during the early and intermediate stages of growth. Before approximately 376 ns, the inhibited system contains fewer sI cages than the uninhibited system. However, after approximately 376 ns, the two curves cross, and the EA-WPU system eventually contains a larger number of canonical sI cages.

This apparent enhancement of sI cage number in the presence of EA-WPU should not be interpreted as an acceleration of hydrate growth, because the number of all detected water cages is markedly lower in the inhibited system throughout the entire simulation. In the system without EA-WPU, the total cage number rapidly increases and reaches approximately 867 cages at 500 ns, whereas the EA-WPU system contains only about 713 cages at the same time. Thus, although the inhibited system eventually possesses more canonical sI cages, it contains substantially fewer total cages. EA-WPU reduces the amount of hydrate-like cage formation but increases the structural selectivity of the cages that survive.

The snapshot in Figure 4(c) provides direct structural support for this interpretation. In the absence of inhibitor, the hydrate growth region contains a dense and highly heterogeneous network of cages, including numerous nonstandard and partial cages represented by colors other than red and blue. At this time, water molecules quickly form cage-like motifs, but many of these motifs do not anneal into the ideal $5^{12}$ and $5^{12}6^{2}$ cages within the simulation time. In contrast, the EA-WPU system exhibits fewer overall cages and a disrupted cage network near the polymer-rich region, but the cages that remain are dominated by the canonical red and blue sI motifs. This behavior is consistent with the general kinetic principle that crystallization rate and crystallinity are often negatively correlated. Rapid growth in the neat system favors the formation of a larger number of kinetically trapped, defective, or amorphous cage structures, whereas the slower growth induced by EA-WPU provides more time for local structural relaxation and selective annealing toward sI-like cages.

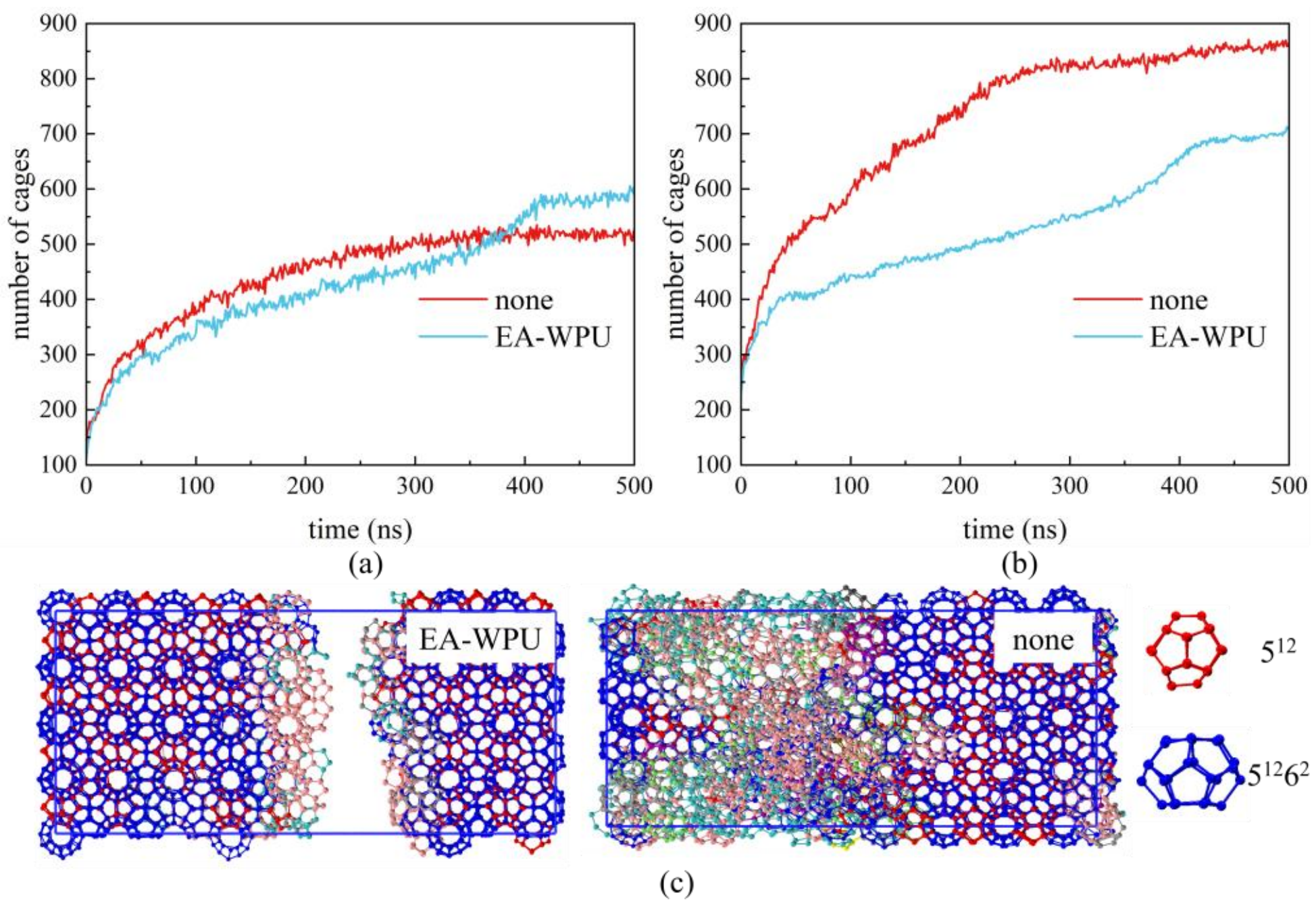

Figure 4. Time evolution of hydrate cage structures. (a) Temporal variation in the total number of $5^{12}$ and $5^{12}6^2$ cages; (b) temporal variation in the total number of water cages; (c) snapshot of the cage structure at 500 ns. For clarity, the $5^{12}$ cages are shown in red, the $5^{12}6^2$ cages in blue, while the remaining colors denote other types of cages or partial cages.

### 3.2. Hydrate Nucleation Pathway

The previous analysis focused on hydrate growth from a pre-existing crystalline phase, where the structural template guides cage formation and the main process is interfacial propagation. Hydrate nucleation, however, is fundamentally more stochastic and pathway-dependent: cages must first emerge from a disordered liquid, pass through transient partial or amorphous intermediates, and eventually assemble into a stable critical nucleus. Therefore, nucleation provides a more stringent test of HyTopo, as it requires resolving not only how many hydrate-like structures form, but also which cage motifs appear, disappear, and interconvert during the earliest stages of self-assembly. This part of the simulation uses the NPT ensemble, with a temperature of 260 K and a pressure of 50 MPa.

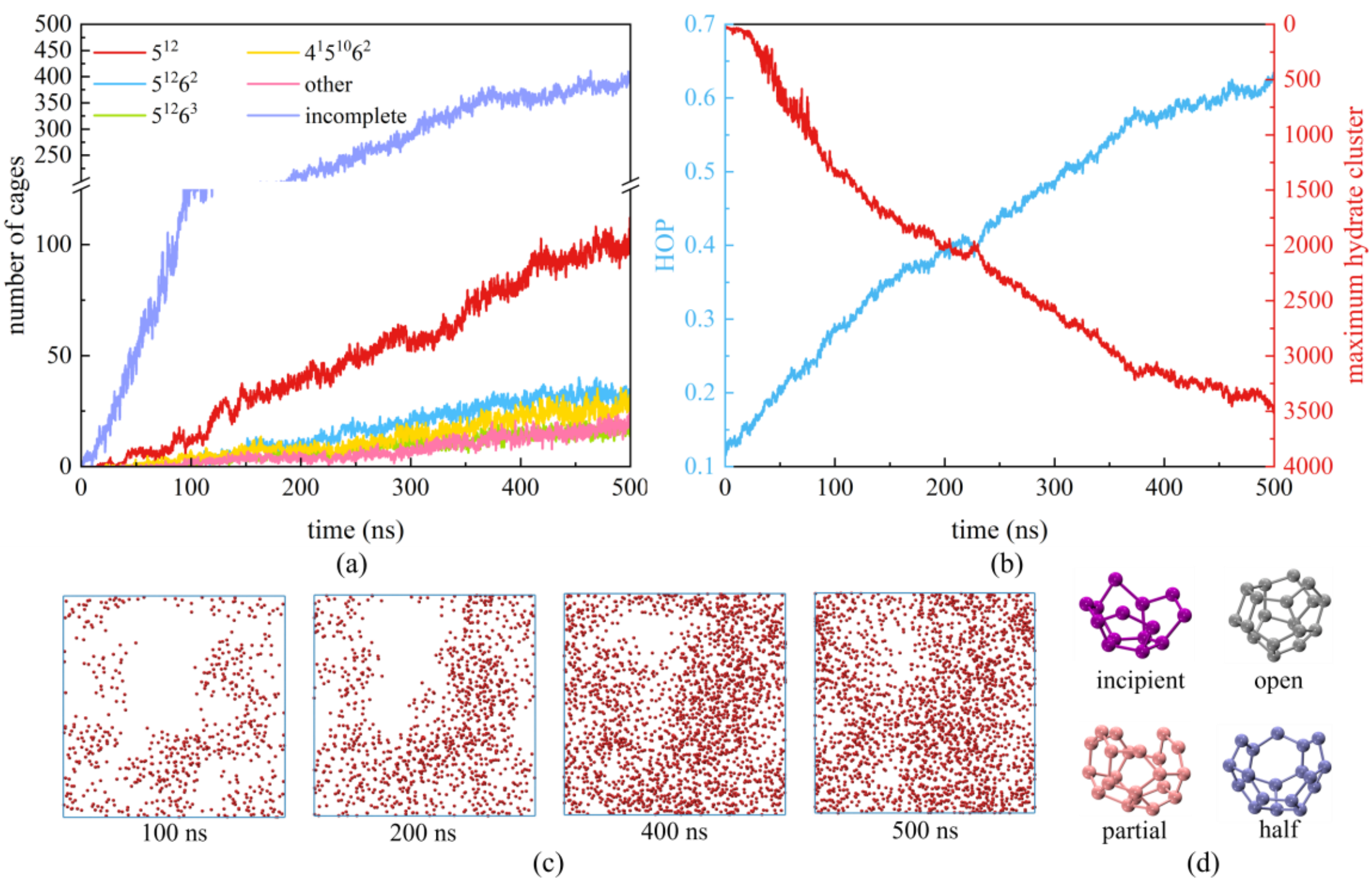

Figure 5. Topology-resolved quantification of methane hydrate nucleation by HyTopo. (a) Time

evolution of cage populations classified by topology. In each frame, cage types with fewer than 10 cages are grouped as "other," so the cage types included in this category may vary over time. "Incomplete" includes half, incipient, open, and partial cages identified by HyTopo.; (b) the global HOP value and the size of the largest hydrate cluster; (c) spatial distributions of oxygen atoms classified as hydrate-like; (d) Schematic diagram of four types of incomplete cages. Incipient cages represent early cage embryos, open cages represent non-closed shells with large boundaries, partial cages represent incomplete structures topologically closest to a target cage, and half-cages represent partially developed shells with substantial but incomplete surface coverage.

Figure 5 presents the HyTopo-based quantitative analysis of methane hydrate nucleation over the 500 ns trajectory. The nucleation process is dominated by incomplete cage structures, whose population increases continuously and reaches approximately 394 by the end of the simulation. In contrast, complete hydrate cages appear more slowly. Among them, the canonical sI cages, $5^{12}$ and $5^{12}6^{2}$ are the most abundant, reaching final populations of approximately 100 and 33, respectively. Nonstandard cages, such as $5^{12}6^{3}$ and $4^{1}5^{10}6^{2}$ , remain below 30 throughout the trajectory. These results indicate that the early stage of hydrate nucleation does not proceed through the direct formation of an ideal sI cage network. Instead, it is characterized by the accumulation of partially closed and topologically defective cage structures, which subsequently serve as a structural reservoir for the formation of complete hydrate cages.

The coupled evolution of HOP and the largest hydrate cluster in Figure 5b further supports a multistep nucleation scenario. The global HOP value increases monotonically to approximately 0.62, reflecting the gradual development of hydrate-like local order. Meanwhile, the largest hydrate cluster grows continuously from a small initial aggregate to approximately 3500 water molecules at the end of the simulation. The spatial distributions in Figure 5c show that, within the first 100 ns, hydrate-like oxygen atoms are sparse and spatially heterogeneous, appearing as isolated clusters rather than a continuous nucleus. As the simulation proceeds, these small hydrate-like clusters become denser and begin to coalesce into a larger connected domain. Even at later stages, however, the spatial distribution remains heterogeneous rather than fully crystalline, consistent with the persistent dominance of incomplete cages over fully closed sI cages. This reflects the fact that the transformation from amorphous hydrate-like structures to well-ordered crystalline hydrate is expected to occur on timescales far beyond conventional molecular dynamics simulations, possibly extending to the millisecond or even second regime.

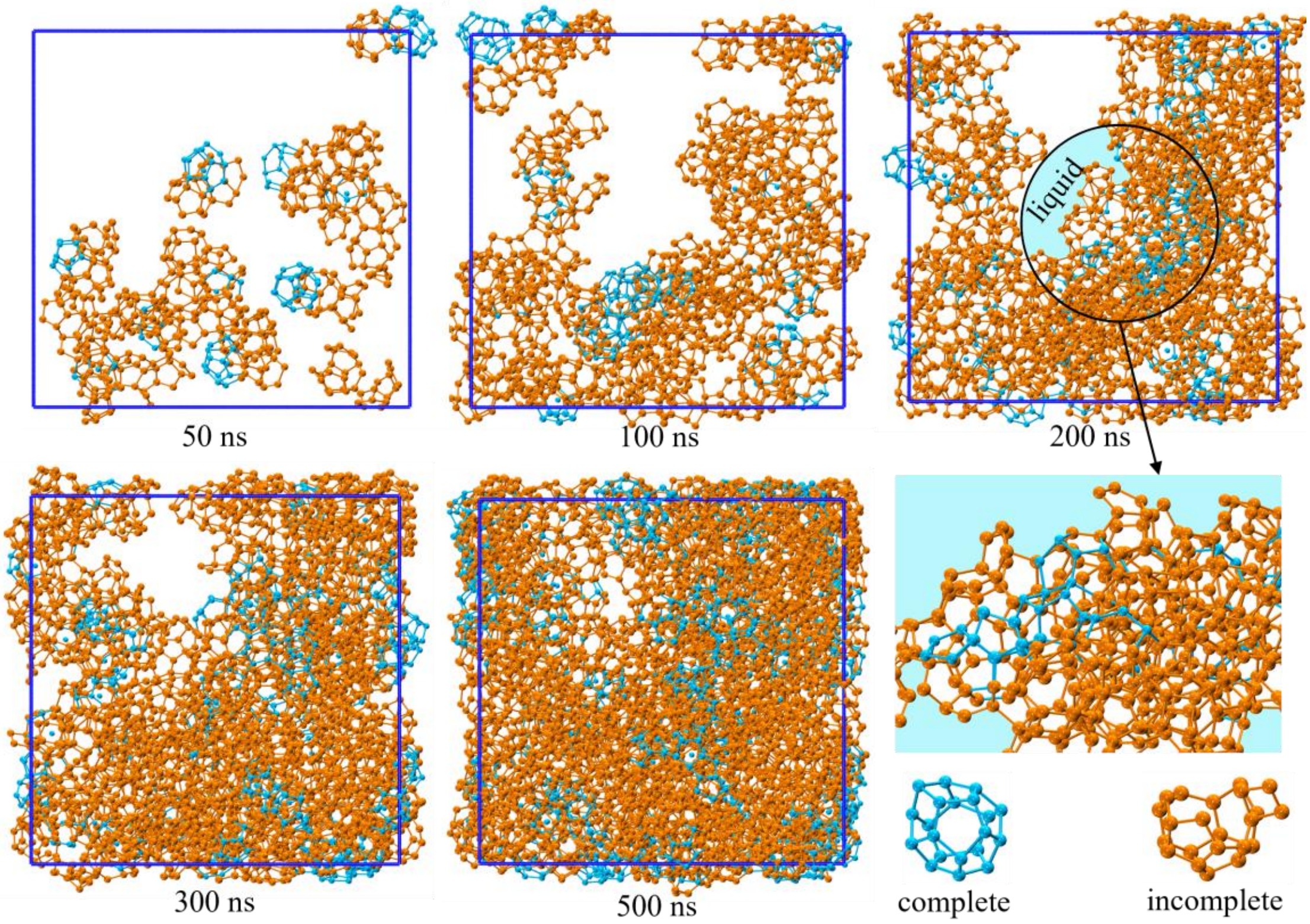

Figure 6. Direct topological visualization of cage networks during hydrate nucleation process

Figure 6 provides a direct topological visualization of the cage network during methane hydrate nucleation, explicitly distinguishing complete and incomplete cages. Consistent with the quantitative statistics in Figure 5, the emerging hydrate phase is dominated by incomplete cage-like structures throughout the trajectory. At the early stage of nucleation, only a few isolated complete cages are observed, and these are embedded within loose clusters of incomplete cages.

As nucleation proceeds, the incomplete cages rapidly proliferate and interconnect, giving rise to an extended amorphous cage network. The spatial arrangement of the cages further reveals the structural role of these incomplete motifs. Rather than appearing randomly, incomplete cages largely surround the complete cages and occupy the interfacial region between the liquid phase and the more ordered cage-rich domains. This topology suggests that incomplete water cages act as transitional states between liquid water and fully closed hydrate cages. Therefore, the formation of regular hydrate cages does not occur directly from the homogeneous liquid phase; instead, it proceeds within an amorphous hydrate-like matrix generated by the accumulation and interconnection of incomplete cage structures. This observation provides direct topological evidence for a multistep nucleation mechanism in which defective cage networks precede and facilitate the emergence of crystalline hydrate motifs.

The spatial organization in Figure 6 suggests that incomplete cages may represent intermediates between liquid water and complete hydrate cages. To test whether these structural relationships correspond to actual temporal pathways, HyTopo tracks individual cages across successive trajectory frames. Each cage is defined by its cage type and set of constituent water-oxygen indices. Cage correspondence between adjacent frames is determined primarily by the Jaccard overlap of these sets, together with guest identity, cage-type continuity, and a limited tolerance for temporarily missing detections. Matched cages are treated as the same evolving structural entity, whereas unmatched detections are recorded as cage formation or disappearance events.

These events are represented as a directed, weighted transition network. Nodes correspond to structural states, including Liquid, incomplete or open cages, intermediate cage motifs, and canonical sI cages. Directed edges denote observed transitions between states; their weights are based on cumulative transition counts and are normalized by the total trajectory duration to obtain transition frequencies. Cage formation and disappearance are linked to the Liquid state, enabling the network to capture both direct cage transformations and the repeated formation and loss of transient structures. These events are represented as a directed, weighted transition network, in which nodes represent structural states, including liquid water and different cage types, and arrows indicate the direction of observed transitions. Edge thickness reflects the transition frequency, with thicker edges indicating more frequent transitions. Transition frequencies are calculated by dividing the cumulative counts of each transition by the total trajectory duration and are expressed in $ns^{-1}$. Cage formation is represented as a transition from the liquid state to a cage state, whereas cage disappearance is represented as a transition from a cage state to the liquid state. Together with transitions between cage types, these connections allow the network to describe cage formation, interconversion, and disappearance. The resulting structural transition network is shown in Figure 7.

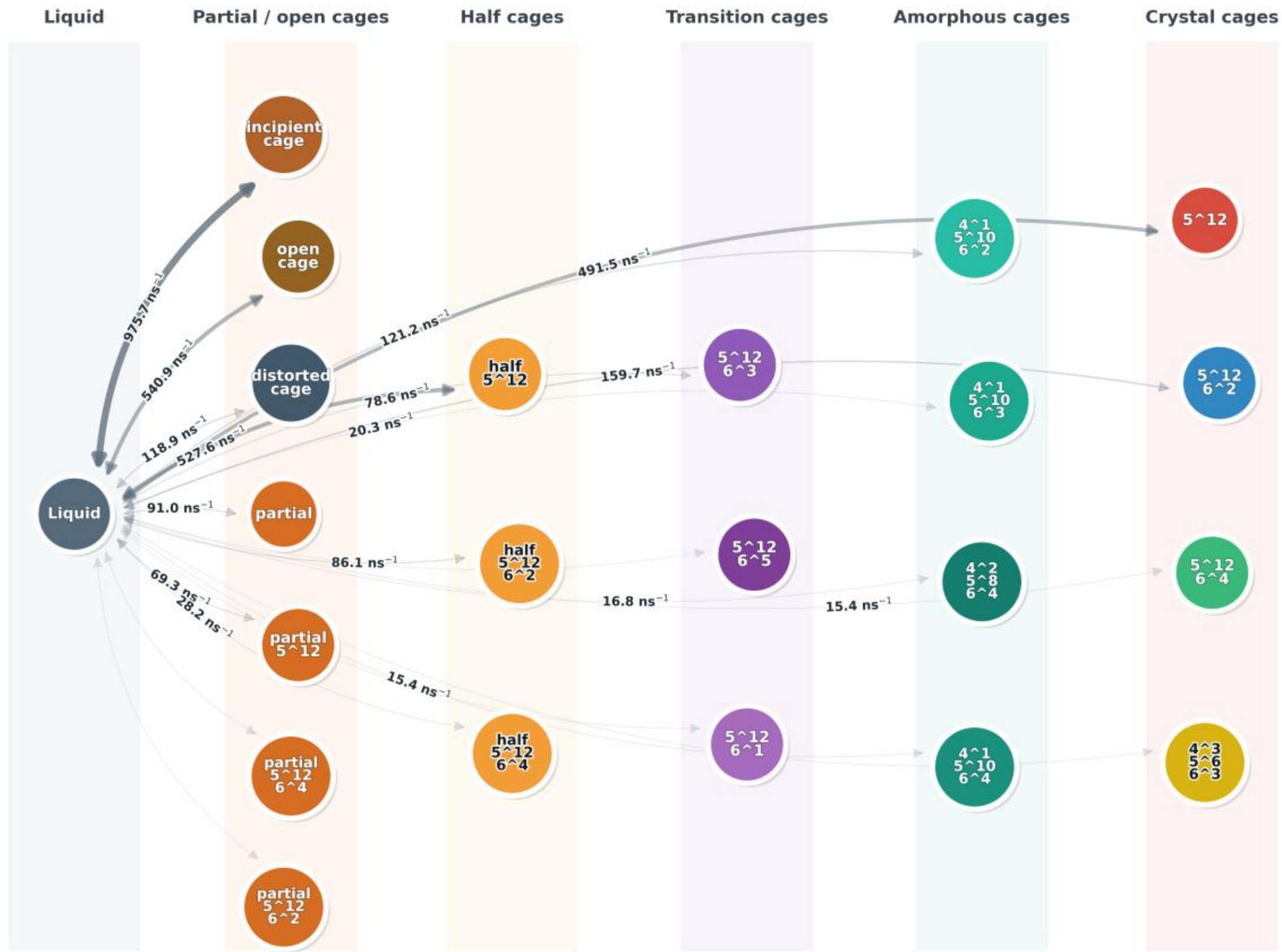

Figure 7. Hydrate cage structural evolution pathway network

The transition network shows that hydrate formation is dominated not by direct, deterministic conversion among a few stable cages, but by frequent exchanges between liquid water and a broad spectrum of incomplete and nonstandard cage motifs. Among these pathways, transitions from liquid water to incipient cages exhibit the highest frequency, reaching 975.7 ns$^{-1}$, indicating that the earliest stage of nucleation is governed by the repeated formation of topologically incomplete water cages. These incipient and defective motifs act as transient intermediates between the disordered liquid and canonical sI cages such as $5^{12}$、$5^{12}6^{2}$. Among the $5^{12}$ cages in the various types of crystal cages, this cage type appeared most frequently, with a frequency of 491.5 ns$^{-1}$.

A notable feature of the network is that direct cage-to-cage transformations are relatively rare. Instead, most structural evolution proceeds through the liquid state as a central exchange node, following a repeated liquid-cage-liquid-cage sequence. This behavior suggests that cage formation during nucleation is highly reversible: incomplete cages are frequently generated from liquid-like configurations, subsequently disrupted, and then reformed into new cage motifs through local hydrogen-bond rearrangements. Therefore, the liquid phase should not be viewed as a passive background, but as an active structural reservoir that continuously supplies and recycles water molecules for cage reconstruction.

The highly branched pathways from the liquid state toward partial, half, distorted, incipient, and nonstandard cages further reveal the topological diversity of the nucleation process. Partial and half cages connect liquid water to more ordered cage types, whereas nonstandard cages provide intermediate routes toward canonical hydrate cages. The relatively dense connectivity among these intermediate states indicates that hydrate nucleation involves continuous topological reorganization, in which defective cages are repeatedly created, modified, and partially annealed. Only a small fraction of these transient structures eventually evolves into complete cages within the amorphous cage network. Thus, the central role of the Liquid node reflects the strong reversibility, short lifetime, and pathway multiplicity characteristic of the early nucleation stage.

### 3.3. Hydrate Decomposition Process

HyTopo was further applied to analyze methane hydrate dissociation under depressurization, one of the most widely used strategies for hydrate gas production. To model pressure-driven decomposition, a vacuum layer was introduced above the sI methane hydrate crystal, and the system was simulated at 277 K in NVT ensemble. By tracking the evolution of standard sI cages, defective intermediates, and partial cage structures, HyTopo enables a topology-resolved description of how the hydrate lattice destabilizes, fragments, and ultimately releases methane during depressurization-induced dissociation.

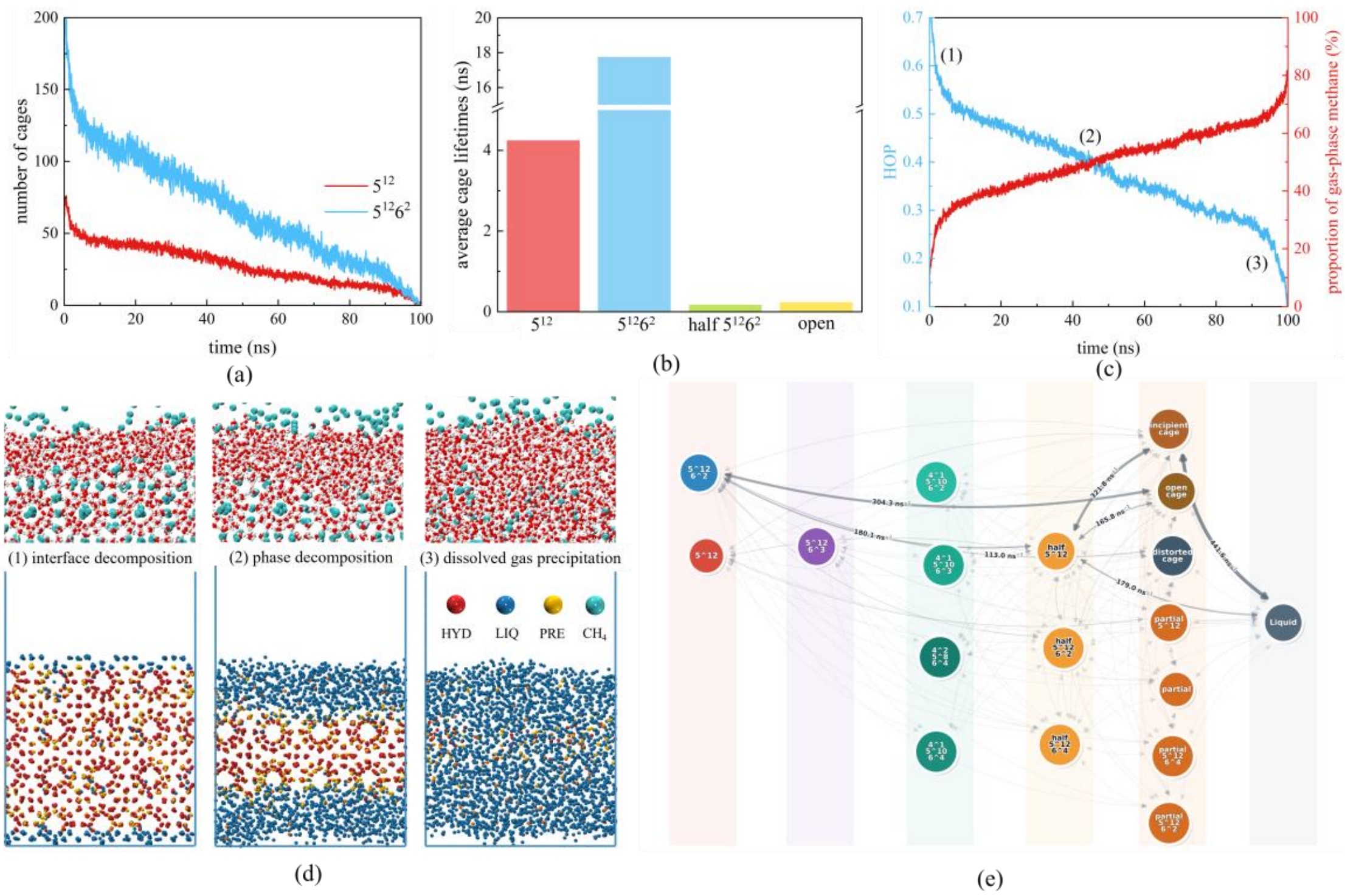

Figure 8. Hydrate depressurization and decomposition process. (a) the number of $5^{12}$ and $5^{12}6^{2}$ cages; (b) the average life time of several types of cages. The lifetime of a cage track was defined as the time interval between its first and last detection around the same guest molecule, allowing continuity based on water-set overlap between consecutive frames. For clarity, cage types with a life time of less than 100 ps are not displayed; (c) HOP parameters and the proportion of gas-phase molecules during the decomposition process; (d) three processes in hydrate decomposition; (e) the transformation path of the cage during the decomposition process.

Figure 8 demonstrates that depressurization-induced methane hydrate dissociation proceeds through three distinct stages rather than by uniform melting of the hydrate lattice. The first stage corresponds to rapid interfacial decomposition and lasts only approximately 5 ns. During this short period, the exposed or weakly connected cage motifs at the hydrate surface are preferentially destabilized, leading to a sharp decrease in both the number of hydrate cages and the HOP value. Methane molecules initially trapped in surface-accessible cages begin to escape, marking the onset of hydrate destabilization. The second stage, extending roughly from 5 to 90 ns, is characterized by slower and more homogeneous bulk decomposition. In this regime, the gas-phase methane fraction increases with an approximately constant slope, indicating a relatively steady release of methane coupled to progressive cage collapse in the hydrate interior. In the final stage, the residual hydrate-like network has largely lost its structural integrity; the rapid increase in gas-phase methane mainly originates from the exsolution of previously dissolved methane rather than from continued decomposition of intact hydrate cages.

The cage lifetime analysis in Figure 8(b) further reveals that hydrate dissociation is strongly topology-dependent. The average lifetime of $5^{12}6^{2}$ cages reaches 17.74 ns, substantially longer than that of $5^{12}$ cages ,which survive for only 4.25 ns on average. This contrast indicates that the larger $5^{12}6^{2}$ cages embedded in the hydrate network act as relatively persistent structural units during depressurization, whereas the smaller $5^{12}$ cages have substantially shorter average lifetimes. In contrast, half cages and open cages exhibit extremely short lifetimes of only 0.17 and 0.23 ns, respectively. These incomplete cages therefore do not behave as stable intermediates that accumulate during dissociation. Instead, they represent short-lived transition structures generated immediately before complete collapse into liquid-like water.

The cage-transition network in Figure 8(e) provides a direct topological description of this dissociation pathway. Most $5^{12}6^{2}$ and $5^{12}$ cages preferentially transform into defective, partial, open, or distorted cage motifs before ultimately losing their cage identity and converting into liquid water. The strong flux from standard cages to defective intermediates, followed by the flux from these intermediates to the liquid state, indicates that hydrate dissociation proceeds through sequential topological rupture: cage deformation, face opening, partial collapse, and final liquefaction. This pathway is fundamentally different from the nucleation process, where liquid water dominates a highly reversible cage-liquid-cage fluctuation network. During dissociation, the transition network is more directional and collapse-driven, reflecting the progressive loss of hydrate topology rather than repeated reconstruction of cage motifs. Thus, HyTopo resolves hydrate decomposition as a heterogeneous, topology-dependent process in which cage stability, methane release, and network fragmentation are tightly coupled.

# 4. Conclusions

We have developed HyTopo as a cage-structure analysis framework for characterizing the structural evolution of gas hydrates in molecular dynamics simulations. The central purpose of HyTopo is not only to determine whether a water molecule is hydrate-like, but also to resolve how water molecules organize into rings, cages, defective intermediates, and connected cage networks. By linking local phase classification with explicit cage topology, HyTopo provides a physically interpretable description of hydrate formation and dissociation beyond conventional scalar order parameters.

The applications presented here demonstrate that this topology-centered description captures mechanistic information that is otherwise difficult to access. In hydrate growth simulations, EA-WPU reduces the total number of water cages throughout the trajectory, confirming its inhibitory effect on hydrate crystal growth. For methane hydrate nucleation, HyTopo reveals a multistep pathway dominated by incomplete cage structures. Rather than directly forming an ideal sI lattice, the system first accumulates partial and defective cages, from which a smaller population of stable $5^{12}$ and $5^{12}6^{2}$ cages subsequently emerges. The early nucleation stage is governed by highly reversible liquid-cage-liquid fluctuations, with liquid water serving as the central structural reservoir. This indicates that the incipient nucleus is better described as an amorphous cage network undergoing continuous topological reorganization than as a small fragment of a perfect hydrate crystal.

For depressurization-induced dissociation, HyTopo reveals a more directional collapse pathway. Hydrate decomposition proceeds through three stages: rapid interfacial decomposition, slower bulk decomposition, and dissolved-gas exsolution after the hydrate-like network has largely collapsed. Cage lifetime analysis shows that $5^{12}6^2$ cages survive much longer than $5^{12}$ cages, whereas half and open cages are extremely short-lived rupture states. The cage-transition network further demonstrates that hydrate dissociation is not the microscopic reverse of nucleation, but a topology-dependent collapse process involving cage deformation, face opening, partial rupture, and final liquefaction.

HyTopo offers a useful analysis tool for hydrate simulations by connecting local phase assignment with explicit cage topology. It avoids the ambiguity of relying on a single order parameter and captures incomplete and nonstandard cages that often dominate nucleation, growth inhibition, and decomposition. HyTopo also enables kinetic analysis of cage evolution, including cage lifetimes and structural transition pathways. These capabilities make HyTopo valuable for studying hydrate behavior under conditions relevant to flow assurance, kinetic inhibition, and gas hydrate resource production, while also providing a foundation for future extensions to other cage-forming molecular materials and complex interfacial environments.

## Data availability

HyTopo are freely available at https://github.com/surfactant80/hytopo/tree/main.

## Declaration of competing interest

The authors declare that they have no known competing financial interests or personal relationships that could have appeared to influence the work reported in this paper.

## Acknowledgments

This research was supported by the National Natural Science Foundation of China (Nos. 42576237, 52206268, 52174255, 42506222), the Shenzhen Science and Technology Program (No. JCYJ20250604144911016), the Natural Science Foundation of Top Talent of SZTU (No. GDRC202314) and the China Postdoctoral Science Foundation General Fund (No. 2024M764183).

## References

(1) Falenty, A.; Hansen, T. C.; Kuhs, W. F. Formation and properties of ice XVI obtained by emptying a type sII clathrate hydrate. *Nature* **2014**, *516* (7530), 231-233. DOI: 10.1038/nature14014.
(2) del Rosso, L.; Celli, M.; Grazzi, F.; Catti, M.; Hansen, T. C.; Fortes, A. D.; Ulivi, L. Cubic ice Ic without stacking defects obtained from ice XVII. *Nature Materials* **2020**, *19* (6), 663-668. DOI: 10.1038/s41563-020-0606-y.
(3) Muromachi, S.; Takeya, S. Discovery of the final primitive Frank-Kasper phase of clathrate hydrates. *Science Advances* **2024**, *10* (30), eadp4384. DOI: doi:10.1126/sciadv.adp4384.
(4) Ibrahim, E.; Lysogorskiy, Y.; Drautz, R.; Piaggi, P. M. Water Phase Diagram from a General-Purpose Atomic Cluster Expansion Potential. *Journal of Chemical Theory and Computation* **2026**, *22* (9), 4758-4766. DOI: 10.1021/acs.jctc.6c00094.
(5) Zysk, F.; Vila Verde, A.; Kaliannan, N. K.; Karhan, K.; Kühne, T. D. Predicting the hydrogen bond strength from water reorientation dynamics at short timescales. *The Journal of Chemical Physics* **2026**, *164* (23). DOI: 10.1063/5.0343765.
(6) Lee, M.; Lee, S. Y.; Kang, M.-H.; Won, T. K.; Kang, S.; Kim, J.; Park, J.; Ahn, D. J. Observing growth and interfacial dynamics of nanocrystalline ice in thin amorphous ice films. *Nature Communications* **2024**, *15* (1), 908. DOI: 10.1038/s41467-024-45234-x.
(7) Tonauer, C. M.; Fidler, L.-R.; Giebelmann, J.; Yamashita, K.; Loerting, T. Nucleation and growth of crystalline ices from amorphous ices. *The Journal of Chemical Physics* **2023**, *158* (14). DOI: 10.1063/5.0143343.
(8) Tian, W.; Wang, C.; Zhou, K. The Dynamic Diversity and Invariance of Ab Initio Water. *Journal of Chemical Theory and Computation* **2024**, *20* (23), 10667-10675. DOI: 10.1021/acs.jctc.4c01191.
(9) Maity, D.; Chakrabarty, S. IceCoder: Identification of Ice Phases in Molecular Simulation Using Variational Autoencoder. *Journal of Chemical Theory and Computation* **2025**, *21* (4), 1916-1928. DOI: 10.1021/acs.jctc.4c01298.
(10) Kanaujiya, R.; Metya, A. K.; Kumar, R.; Patra, T. K. Nanobubble size controls gas hydrate nucleation in supercooled water. *Physical Chemistry Chemical Physics* **2026**, *28* (22), 13469-13476. DOI: 10.1039/d5cp04275e.
(11) Fiedler, F.; Vinš, V.; Jäger, A.; Span, R. Modification of the van der Waals and Platteeuw model for gas hydrates considering multiple cage occupancy. *The Journal of Chemical Physics* **2024**, *160* (9). DOI: 10.1063/5.0189555.

(12) Fang, Y.; Li, X.; Wang, H.; Pan, Y.; Guo, M.; Jin, D. Molecular Simulation Study on the Gas Selectivity of the Mixed Carbon Dioxide/Methane Hydrates Confined in Nanoporous Carbon. *The Journal of Physical Chemistry B* **2026**, *130* (18), 4838-4848. DOI: 10.1021/acs.jpcb.6c00981.
(13) Yoshida, K.; Uematsu, K.; Noguchi, N. Pressure-Induced Amorphization of Methane Hydrate: An Ab Initio Molecular Dynamics Study of Guest Molecule Solvation. *The Journal of Physical Chemistry B* **2026**, *130* (20), 5202-5210. DOI: 10.1021/acs.jpcb.6c00818.
(14) Rao, S.; Li, Z.; Lu, H.; Deng, Y. A review of gas hydrate formation characteristics at interfaces. *Fuel* **2025**, *392*, 134863. DOI: 10.1016/j.fuel.2025.134863.
(15) Wang, H.; Li, Y.; Huang, L.; Liu, T.; Liu, W.; Wu, P.; Song, Y. A pore-scale study on microstructure and permeability evolution of hydrate-bearing sediment during dissociation by depressurization. *Fuel* **2024**, *358*, 130124. DOI: 10.1016/j.fuel.2023.130124.
(16) E Dendy, S. Fundamental principles and applications of natural gas hydrates. *Nature* **2003**, *426* (6964), 353-363. DOI: 10.1038/nature02135.
(17) Chen, Z.; Song, X.; Deng, Y.; Li, Z.; Huang, S.; Tang, D.; Liu, Q. Molecular dynamics evaluation of modified surfactants as dual-function inhibitors for hydrate agglomeration and asphalt deposition. *Fuel* **2027**, *427*, 139820. DOI: 10.1016/j.fuel.2026.139820.
(18) Wang, L.; Gu, L.; Zhan, L.; Cai, W.; Lu, H. Long-term safety of continental slope carbon storage in the South China Sea: 600,000-year simulation of the effect of hydrate cap and density pump coupling mechanism. *Energy* **2026**, *355*, 141131. DOI: 10.1016/j.energy.2026.141131.
(19) Kou, X.; Zhang, H.; Chen, Z.-Y.; Wang, Y.; Li, X.-S. Dynamic mechanisms of gas hydrate evolution in hydrophobic sediments: Wall-climbing and secondary formation. *Energy* **2026**, *353*, 141013. DOI: 10.1016/j.energy.2026.141013.
(20) Xie, Y.; Feng, J.; Xu, L.; Chen, X.; Wang, J.; Peng, F.; Qi, Z.; Wang, B.; Wang, Y.; Zhang, S.; et al. Optimizing the depressurization rate to balance CH4 hydrate dissociation and CH4 leakage in unconfined marine hydrate reservoirs. *Fuel* **2027**, *427*, 139559. DOI: 10.1016/j.fuel.2026.139559.
(21) Wu, P.; Liu, S.; Pinkert, S.; Wang, Y.; Huang, L.; Song, Y.; Li, Y. Undrained deformation of hydrate-bearing clayey silt could generate submarine landslides. *Journal of Ocean Engineering and Science* **2026**. DOI: 10.1016/j.joes.2026.04.009.
(22) Chen, Z.; Farhadian, A.; Sadeh, E.; Chen, C.; Striolo, A. Counterion-dependent promotion of methane hydrate formation by anionic biosurfactants: A molecular perspective for efficient gas storage. *Chemical Engineering Journal* **2025**, *526*, 170846. DOI: doi.org/10.1016/j.cej.2025.170846.
(23) Mathews, S.; Zhu, X.; Guerra, A.; Servio, P.; Rey, A. Atomistic Modeling of Methane and Carbon Dioxide Structure I Gas Hydrates under Pressure: Guest Effects and Properties. *Journal of Chemical Theory and Computation* **2026**, *22* (6), 3114-3124. DOI: 10.1021/acs.jctc.5c01868.
(24) Luo, T.; Ma, C.; Chen, Z.; Zhu, Y.; Shangguan, Y.; Yang, D.; Yang, W.; Song, Y. Enhanced Seafloor CO2 Sequestration via Hydrate Formation in Fractures: Experimental Investigation of Kinetics and Morphological Characteristics. *Energy & Fuels* **2025**, *39* (33), 15807-15821. DOI: 10.1021/acs.energyfuels.5c02498.
(25) Ait Blal, A.; Lu, P.; Al-Atrach, J.; Guillet-Nicolas, R.; Castellani, B.; Valtchev, V. A mechanistic study of CO2 gas hydrate formation in a mesoporous zeolite. *Nature Communications* **2025**, *17* (1), 310. DOI: 10.1038/s41467-025-67019-6.
(26) Chen, L.; Ting, V. P.; Zhang, Y.; Coventry, J.; Rahbari, A.; Yin, Z.; Wang, F.; Tian, M.; Rochat, S.; Zhang, Z.; et al. Hydrate-based H2 storage with porous materials as heterogeneous promoters: state of the art and challenges. *Journal of Materials Chemistry A* **2025**, *13* (33), 27001-27049. DOI:

10.1039/d5ta04503g.
(27) Cathalot, C.; Rinnert, E.; Scalabrin, C.; Fandino, O.; Giunta, T.; Ondreas, H.; Rouxel, O.; Rabouille, C.; Dumoulin, J. P.; Bombled, B.; et al. Large CO2 seeps and hydrate field on the seafloor offshore Mayotte Island. *Nature Geoscience* **2026**. DOI: 10.1038/s41561-026-02004-2.
(28) Ma, C.; Luo, T.; Zhang, C.; Zhang, Y.; Zhu, X.; Han, T.; Li, Y.; Song, Y.; Yang, W. Laboratory experiments and empirical normalized model of water-phase permeability in CO2 hydrate-bearing silty-clayey sediments. *Fuel* **2027**, *428*, 140341. DOI: 10.1016/j.fuel.2026.140341.
(29) Tanaka, H.; Matsumoto, M.; Yagasaki, T.; Takeuchi, M.; Mori, Y.; Kono, T. Phase behaviour of liquid CO2 with an impurity of water: influence of CO2 hydrate. *Physical Chemistry Chemical Physics* **2026**, *28* (24), 15002-15014. DOI: 10.1039/d6cp01072e.
(30) Grabowska, J.; Kuffel, A.; Zielkiewicz, J. Revealing the Frank–Evans “Iceberg” Structures within the Solvation Layer around Hydrophobic Solutes. *The Journal of Physical Chemistry B* **2021**, *125* (6), 1611-1617. DOI: 10.1021/acs.jpcb.0c09489.
(31) Galamba, N. Water’s Structure around Hydrophobic Solutes and the Iceberg Model. *The Journal of Physical Chemistry B* **2013**, *117* (7), 2153-2159. DOI: 10.1021/jp310649n.
(32) Chen, Z.; Farhadian, A.; Sadeh, E.; Chen, C. Micellization effects in surfactant-enhanced gas hydrate formation for efficient solidified methane storage. *Energy* **2025**, *332*, 137088. DOI: 10.1016/j.energy.2025.137088.
(33) Liu, X.-Y.; Yan, X.-Y.; Liu, Y.; Qu, H.; Wang, Y.; Wang, J.; Guo, Q.-Y.; Lei, H.; Li, X.-H.; Bian, F.; et al. Self-assembled soft alloy with Frank–Kasper phases beyond metals. *Nature Materials* **2024**, *23* (4), 570-576. DOI: 10.1038/s41563-023-01796-7.
(34) Lee, S.; Bluemle, M. J.; Bates, F. S. Discovery of a Frank-Kasper σ Phase in Sphere-Forming Block Copolymer Melts. *Science* **2010**, *330* (6002), 349-353. DOI: doi:10.1126/science.1195552.
(35) Bissuel, D.; Orveillon, L.; Arrondeau, B.; Almeida de Mendonça, J. P.; Daubin, J.; Piazza, I.; Uhrin, M.; Polack, É.; Kandy, A. K. A.; Martin-Calle, D.; et al. Reproducible Container Solutions for Codes and Workflows in Materials Science. *Advanced Engineering Materials n/a* (n/a), e202503185. DOI: 10.1002/adem.202503185.
(36) Kirov, M. V. Topological identification of multi-cage water clusters. *The Journal of Chemical Physics* **2026**, *164* (18). DOI: 10.1063/5.0324777.
(37) Berni, S.; Espert, S.; Poręba, T.; Di Cataldo, S.; Gaal, R.; Tobie, G.; Le Menn, E.; Hansen, T. C.; Bini, R.; Bove, L. E. Discovery of a low-density filled-ice phase in nitrogen hydrate at high pressure. *Communications Chemistry* **2026**. DOI: 10.1038/s42004-026-02030-6.
(38) Yuan, J.; Li, M.; Wang, H. High-throughput computational screening of porous materials for CO2 removal from Fischer–Tropsch synthesis. *Materials Genome Engineering Advances* **2025**, *3* (3), e70023. DOI: 10.1002/mgea.70023.
(39) Seol, J. Water-miscible amine-hydrocarbon hybrid strategy for methane storage via hydrates: Simultaneous enhancement of kinetics, gas uptake, and stability. *Chemical Engineering Journal* **2026**, 179154. DOI: 10.1016/j.cej.2026.179154.
(40) Ke, W.; Svartaas, T. M.; Chen, D. A review of gas hydrate nucleation theories and growth models. *Journal of Natural Gas Science and Engineering* **2019**, *61*, 169-196. DOI: 10.1016/j.jngse.2018.10.021.
(41) Dong, Z.; Liu, J.; Cao, J.; Wu, E.; Chi, F.; Liu, S. Unveiling hydrate nucleation mechanisms: gas concentration thresholds and memory effect origins. *Journal of Molecular Liquids* **2025**, *436*, 128213. DOI: 10.1016/j.molliq.2025.128213.
(42) Guo, G.-J.; Zhang, Z. Open questions on methane hydrate nucleation. *Communications Chemistry*

**2021**, *4* (1), 102. DOI: 10.1038/s42004-021-00539-6.
(43) Li, Y.; Zeng, H.; Zhang, H. Atomistic simulations of nucleation and growth of CaCO3 with the influence of inhibitors: A review. *Materials Genome Engineering Advances* **2023**, *1* (1), e4. DOI: 10.1002/mgea.4.
(44) Wang, S.; Wang, J.; Xue, C.; Yang, X.; Tian, G.; Su, H.; Miao, Y.; Li, Q.; Li, X. Unexpected nucleation mechanism of T1 precipitates by Eshelby inclusion with unstable stacking faults. *Materials Genome Engineering Advances* **2024**, *2* (2), e33. DOI: 10.1002/mgea.33.
(45) Finney, A. R.; Salvalaglio, M. Molecular simulation approaches to study crystal nucleation from solutions: Theoretical considerations and computational challenges. *WIREs Computational Molecular Science* **2024**, *14* (1), e1697. DOI: 10.1002/wcms.1697.
(46) Wang, Z.; Yuan, Z.; Cheng, M.; Huang, X.; Liu, K.; Wang, Y.; Sun, H.; Liao, L.; Xu, Z.; Chen, J.; et al. Molecularly resolved mapping of heterogeneous ice nucleation and crystallization pathways using in-situ cryo-TEM. *Nature Communications* **2025**, *16* (1), 7349. DOI: 10.1038/s41467-025-62900-w.
(47) Liu, J.; Dong, Z.; Wang, S.; Yan, Y.; Gao, Z.; Zhang, Y. Molecular dynamics simulation of methane hydrate formation in the presence of nanobubbles. *Chemical Engineering Science* **2026**, *324*, 123325. DOI: 10.1016/j.ces.2026.123325.
(48) Walsh, M. R.; Koh, C. A.; Sloan, E. D.; Sum, A. K.; Wu, D. T. Microsecond Simulations of Spontaneous Methane Hydrate Nucleation and Growth. *Science* **2009**, *326* (5956), 1095-1098. DOI: 10.1126/science.1174010.
(49) He, Z.; Gupta, K. M.; Linga, P.; Jiang, J. Molecular Insights into the Nucleation and Growth of CH4 and CO2 Mixed Hydrates from Microsecond Simulations. *The Journal of Physical Chemistry C* **2016**, *120* (44), 25225-25236. DOI: 10.1021/acs.jpcc.6b07780.
(50) Hu, W.; Chen, C.; Sun, J.; Zhang, N.; Zhao, J.; Liu, Y.; Ling, Z.; Li, W.; Liu, W.; Song, Y. Three-body aggregation of guest molecules as a key step in methane hydrate nucleation and growth. *Communications Chemistry* **2022**, *5* (1), 33. DOI: 10.1038/s42004-022-00652-0.
(51) Nguyen, A. H.; Molinero, V. Identification of Clathrate Hydrates, Hexagonal Ice, Cubic Ice, and Liquid Water in Simulations: the CHILL+ Algorithm. *The Journal of Physical Chemistry B* **2015**, *119* (29), 9369-9376. DOI: 10.1021/jp510289t.
(52) Mahmoudinobar, F.; Dias, C. L. GRADE: A code to determine clathrate hydrate structures. *Computer Physics Communications* **2019**, *244*, 385-391. DOI: 10.1016/j.cpc.2019.06.004.
(53) Liu, Y.; Xu, K.; Xu, Y.; Liu, J.; Wu, J.; Zhang, Z. HTR: An ultra-high speed algorithm for cage recognition of clathrate hydrates. *Nanotechnology Reviews* **2022**, *11* (1), 699-711. DOI: 10.1515/ntrev-2022-0044.
(54) Jacobson, L. C.; Hujo, W.; Molinero, V. Amorphous Precursors in the Nucleation of Clathrate Hydrates. *Journal of the American Chemical Society* **2010**, *132* (33), 11806-11811. DOI: 10.1021/ja1051445.
(55) Jacobson, L. C.; Hujo, W.; Molinero, V. Nucleation Pathways of Clathrate Hydrates: Effect of Guest Size and Solubility. *Journal of Physical Chemistry B* **2010**, *114* (43), 13796-13807. DOI: 10.1021/jp107269q.
(56) Bagherzadeh, S. A.; Alavi, S.; Ripmeester, J. A.; Englezos, P. Why ice-binding type I antifreeze protein acts as a gas hydrate crystal inhibitor. *Phys Chem Chem Phys* **2015**, *17* (15), 9984-9990. DOI: 10.1039/c4cp05003g.
(57) Noé, F.; Fischer, S. Transition networks for modeling the kinetics of conformational change in macromolecules. *Current Opinion in Structural Biology* **2008**, *18* (2), 154-162. DOI:

10.1016/j.sbi.2008.01.008.
(58) Chodera, J. D.; Noé, F. Markov state models of biomolecular conformational dynamics. *Current Opinion in Structural Biology* **2014**, *25*, 135-144. DOI: 10.1016/j.sbi.2014.04.002.
(59) Chen, Z.; Farhadian, A.; Naeiji, P.; Martyushev, D. A.; Chen, C. Molecular-Level insights into kinetic and agglomeration inhibition mechanisms of structure I and II gas hydrate formation. *Chemical Engineering Journal* **2025**, *511*, 162194. DOI: 10.1016/j.cej.2025.162194.
(60) Farhadian, A.; Gnezdilov, D.; Semenov, M.; Qaid, M. A. A. A.; Varfolomeev, M. Relationship Between the End-Cap Structure of Polyurethanes and Their Kinetic Hydrate Inhibition Activity: Effective Hydrate Prevention at Low Concentrations. *Energy & Fuels* **2024**, *38* (17), 16089-16100. DOI: 10.1021/acs.energyfuels.4c02628.